\documentclass[notitlepage,nofootinbib,preprintnumbers,amssymb,superscriptaddress]{revtex4-1}
\usepackage{amsfonts,amssymb,mathtools,graphicx,color,bm}
\definecolor{ultramarine}{rgb}{0.07, 0.04, 0.56}
\definecolor{cadmiumgreen}{rgb}{0.0, 0.42, 0.24}
\definecolor{indigo(dye)}{rgb}{0.0, 0.25, 0.42}
\usepackage[linktocpage=true]{hyperref}
\hypersetup{
colorlinks=true,
citecolor=ultramarine,
linkcolor=cadmiumgreen,
urlcolor=indigo(dye),
}

\usepackage{autobreak}
\newcommand{\fr}[2]{\frac{#1}{#2}}
\newcommand{\pa}{\partial}
\newcommand{\ti}{\tilde}
\newcommand{\na}{\nabla}
\newcommand{\bra}[1]{\left( #1 \right)}  
\newcommand{\brb}[1]{\left[ #1 \right]}  
\newcommand{\brc}[1]{\left\{ #1 \right\}}  
\newcommand{\be}{\begin{equation}}  
\newcommand{\ee}{\end{equation}}
\newcommand{\bem}{\begin{bmatrix}}
\newcommand{\eem}{\end{bmatrix}}
\newcommand{\Mpl}{M_{\rm Pl}}

\newcommand{\ga}{\gamma}

\newcommand{\la}{\lambda}
\newcommand{\si}{\sigma}

\newcommand{\mn}{{\mu \nu}}
\newcommand{\mE}{\mathcal{E}}

\newcommand{\mG}{\mathcal{G}}

\newcommand{\mO}{\mathcal{O}}

\begin{document}

\preprint{KOBE-COSMO-19-18}

\title{Ghost-free resummation of gravitational interactions of a 2-form gauge field}

\author{Kazufumi Takahashi}
\affiliation{Department of Physics, Kobe University, Kobe 657-8501, Japan}

\author{Daisuke Yoshida}
\affiliation{Department of Physics, Kobe University, Kobe 657-8501, Japan}

\begin{abstract}
We find that the recently-proposed ghost-free interaction of a 2-form gauge field in four dimensions, which contains derivative couplings in a nonperturbative manner, can be regarded as a resummation of ghostly interaction terms.
We investigate the higher derivative structure of this model in a minisuperspace description and demonstrate that the higher derivative terms can be removed by taking appropriate combinations of the Euler-Lagrange equations, while a truncation at a finite order spoils this structure. 
We also show that this nature is peculiar to four dimensions.
\end{abstract}

\maketitle

\section{Introduction}\label{sec:intro}

Paradoxically, one of the best ways to understand general relativity is to study modified theories of gravity. 
They could highlight peculiar aspects of general relativity and also serve as a reference of comparison when testing gravity.
One can modify general relativity by violating at least one of the assumptions of Lovelock's theorem~\cite{Lovelock:1971yv}, e.g., by adding some field(s) on top of the metric.
When it comes to constructing an action for such theories, there is a guiding principle called the theorem of Ostrogradsky~\cite{Woodard:2015zca}.
The theorem states that extra degrees of freedom~(DOFs) associated with higher derivatives always result in an unbounded Hamiltonian (called てぇOstrogradsky ghost).
A caveat is that the theorem does not necessarily prohibit the appearance of higher derivatives in the action. 
The problem of the Ostrogradsky ghost can be circumvented if the Euler-Lagrange (EL) equations are of second order or if they can be recomposed into second-order differential equations.
Nevertheless, the theorem of Ostrogradsky severely constrains the higher derivative structure of the action~\cite{Motohashi:2014opa,Langlois:2015cwa,Motohashi:2016ftl,Klein:2016aiq,Motohashi:2017eya,Motohashi:2018pxg}.

In the case of single-field scalar-tensor theories (i.e., whose ingredients are a metric and a scalar field) in four dimensions, we know the most general class of theories that yield second-order EL equations, which is now referred to as the Horndeski class~\cite{Horndeski:1974wa,Deffayet:2011gz,Kobayashi:2011nu}.
There are also yet broader classes of scalar-tensor theories without extra DOFs~\cite{Gleyzes:2014dya,Langlois:2015cwa,Crisostomi:2016czh,BenAchour:2016fzp,Takahashi:2017pje,Langlois:2018jdg}.
Those classes have higher-order EL equations, but they can be reduced to second-order differential equations by taking their linear combinations in an appropriate manner.
It should also be noted that there have been similar advances in vector-tensor theories~\cite{Horndeski:1976gi,Heisenberg:2014rta,Kimura:2016rzw}.
Despite such progress in scalar-tensor and vector-tensor theories, regarding nontrivial interactions between a 2-form gauge field and gravity, we know only a few examples that can avoid the appearance of the Ostrogradsky ghost.
A study of such interactions is important because a 2-form gauge field appears universally in low-energy limits of string theories.
An example of nontrivial interaction was obtained in the context of $p$-form Galileons~\cite{Deffayet:2010zh} in a seven-dimensional spacetime. However, similar constructions do not work in four dimensions.
The authors of Ref.~\cite{Heisenberg:2019akx} investigated possible interactions in four dimensions to find that any interaction with a finite number of derivative couplings yields higher-order EL equations.
This fact underlines the difficulty of constructing healthy derivative interactions of a 2-form gauge field.
On the contrary, an example of nontrivial coupling to gravity in four dimensions was found in Ref.~\cite{Yoshida:2019dxu} by exploring an electromagnetic duality of a healthy scalar-tensor theory. 
A remarkable feature of this model is that it consists of nonperturbative derivative couplings.
Nevertheless, due to the ghost-free nature of its dual scalar-tensor theory, the 2-form model should avoid the problem of the Ostrogradsky ghost.
Also, this model provides a new nonsingular early universe model where an inflationary universe is created from a completely periodic cyclic universe, called a time crystal universe, as investigated in Ref.~\cite{Yoshida:2019pgn}. 

Given the fact that the 2-form model having nonperturbative derivative couplings~\cite{Yoshida:2019dxu} is free of ghost, it should be understood as a nonperturbative ghost-free UV completion of ghostly derivative couplings at low-energy scales.
The main aim of this paper is to demonstrate how the ghost disappears by this completion.
To this end, we move to a minisuperspace description and show that the model has no extra DOFs, while a truncation of the Lagrangian at a finite order results in an unwanted ghostly DOF.
We also show that the model described by the same Lagrangian but defined in five or higher dimensions yields extra DOFs, and thus the absence of extra DOFs is peculiar to four dimensions.

The rest of this paper is organized as follows.
In \S \ref{sec:model}, we provide a brief review of the model proposed in Ref.~\cite{Yoshida:2019dxu}.
In \S \ref{sec:4D}, we develop a minisuperspace description of the model in four dimensions and demonstrate the minisuperspace Lagrangian is free of extra DOFs associated with higher derivatives.
In \S \ref{sec:truncation}, we consider a truncation of the model at the linear order in the curvature tensor to see that there does exist a ghostly DOF.
In \S \ref{sec:higherD}, we study the model in five or higher dimensions and discuss the difference from the case of four dimensions.
Finally, in \S \ref{sec:conc}, we draw our conclusions and present succinct mechanical models which help us to capture the essence of the results in \S \ref{sec:4D} to \S \ref{sec:higherD}.

\section{The Model}\label{sec:model}
In this paper, we focus on the following model defined in four dimensions~\cite{Yoshida:2019dxu}:
    \be
    L=\sqrt{-g}\brb{\fr{\Mpl^2}{2}(R-2\Lambda)-\fr{1}{12}\fr{1}{\det\mG}\mG^{\mu\alpha}\mG^{\nu\beta}\mG^{\la\ga}H_{\mu\nu\la}H_{\alpha\beta\ga}}, \label{Lag}
    \ee
where $H_{\mu\nu\la}\coloneqq 3\na_{[\mu}B_{\nu\la]}$ is the field strength of a 2-form gauge field~$B_\mn$, and $\mG_\mn$ denotes a linear combination of the metric~$g_\mn$ and the Einstein tensor~$G_\mn$,
    \be
    \mG_\mn\coloneqq \alpha g_\mn+\beta G_\mn,
    \ee
with $\alpha$ and $\beta$ being nonvanishing constants.
We assume $\det\mG\coloneqq \det(\alpha\delta^\mu_\nu+\beta G^\mu_\nu)$ is nonvanishing so that the Lagrangian is well defined.
Note that $\det\mG$ transforms as a scalar under general coordinate transformations and it can be expressed as
    \begin{align}
    \alpha^{-4}\det\mG&=1+\sum_{n=1}^4\fr{1}{n!}\bra{\fr{\beta}{\alpha}}^n\delta^{\mu_1\cdots\mu_n}_{\nu_1\cdots\nu_n}G^{\nu_1}_{\mu_1}\cdots G^{\nu_n}_{\mu_n} \nonumber \\
    &=1-\fr{\beta}{\alpha}R-\fr{1}{2}\bra{\fr{\beta}{\alpha}}^2\bra{\langle R^2 \rangle -R^2}
    +\fr{1}{6}\bra{\fr{\beta}{\alpha}}^3\bra{2 \langle R^3 \rangle-\fr{1}{2}R^3} \nonumber \\
    &\quad -\fr{1}{24}\bra{\fr{\beta}{\alpha}}^4\brb{6 \langle R^4 \rangle - 3\langle R^2 \rangle^2-4R \langle R^3 \rangle+3R^2 \langle R^2 \rangle-\fr{1}{2}R^4},
    \end{align}
in four dimensions, with $ \langle R^n \rangle\coloneqq R^{\mu_1}_{\mu_2}R^{\mu_2}_{\mu_3}\cdots R^{\mu_n}_{\mu_1}$.
A remarkable feature of the Lagrangian~\eqref{Lag} is that the curvature tensor (and thus the second derivative of the metric) appears in the denominator.
It is also interesting to note that the model is dual to the following scalar-tensor theory within the Horndeski class~\cite{Yoshida:2019dxu}:
    \be
    \ti{L}=\sqrt{-g}\brb{\fr{\Mpl^2}{2}(R-2\Lambda)-\fr{1}{2}\mG^\mn\na_\mu\phi\na_\nu\phi},
    \ee
and thus should be free of the Ostrogradsky ghost.

In the low-energy limit~$(\beta \times \text{curvature}) \ll 1$, the coupling with the Einstein tensor is written as the standard kinetic term of a 2-form gauge field plus derivative corrections,
\begin{align}
 L=\sqrt{-g} \biggl\{ &\fr{\Mpl^2}{2}(R-2\Lambda)-\fr{1}{12\alpha}H^{\mu\nu\la}H_{\mu\nu\la}  - \frac{1}{24 \alpha} \left(\frac{\beta}{\alpha}\right)  \left( 6 R^{\mu \nu} - R g^{\mu \nu} \right) H_{\mu\alpha\beta}H_{\nu}{}^{\alpha\beta}  \notag\\
 &- \frac{1}{48 \alpha} \left( \frac{\beta}{\alpha} \right)^2 \left[ 12 R^{\mu\la} R^{\nu\sigma} + g^{\mu\la} g^{\nu\sigma} (2 \langle R^2\rangle - R^2) \right] H_{\mu\nu\alpha}H_{\la\sigma}{}^ {\alpha} + \mO(\beta^{3}) \biggr\}.\label{expandedL}
\end{align}
Based on the discussion in Ref.~\cite{Heisenberg:2019akx}, each of the derivative corrections yields higher derivative terms in the EL equations. Thus, our model~\eqref{Lag} can be understood as a ghost-free UV completion of ghostly derivative interactions~\eqref{expandedL}.
In the following sections, we clarify how the completion removes the ghost in a minisuperspace description.

\section{Absence of the ghost in four dimensions}\label{sec:4D}

We impose the following ansatz for the 2-form field:
\begin{align}
 \fr{1}{2}B_\mn dx^\mu \wedge dx^\nu=v(t) dx \wedge dy 
+ \frac{h}{3} \left( x dy \wedge dz  + y dz \wedge dx + z dx \wedge dy \right) , \label{Bmn4D}
\end{align}
with $h$ being constant.
For this configuration, the nontrivial components of the field strength are $H_{012}=\dot{v}$ and $H_{123}=h$.
Correspondingly, the metric is assumed to be of the following form since a nonvanishing $\dot{v}$ makes the spacetime anisotropic:
    \be
    g_\mn dx^\mu dx^\nu=-N^2(t)dt^2+a^2(t)\brb{e^{2\si(t)}(dx^2+dy^2)+e^{-4\si(t)}dz^2}. \label{gmn4D}
    \ee
In the case where $\dot{v}=0$ and thus the spacetime is isotropic (i.e., $\si=0$), the existence of $h$ accommodates nontrivial cosmology~\cite{Yoshida:2019pgn}.
However, for simplicity, we hereafter set $h=0$ as $h$ does not contribute to higher derivative terms.
Noting that the Einstein tensor computed from the metric~\eqref{gmn4D} is diagonal, we can simplify the coupling between the 2-form and the curvature tensor appearing in Eq.~\eqref{Lag} as
    \be
    -\fr{1}{12}\fr{1}{\det\mG}\mG^{\mu\alpha}\mG^{\nu\beta}\mG^{\la\ga}H_{\mu\nu\la}H_{\alpha\beta\ga}
    =-\fr{\dot{v}^2}{2}\fr{\mG^{00}\mG^{11}\mG^{22}}{\mG^0_0\mG^1_1\mG^2_2\mG^3_3}
    =-\fr{\dot{v}^2}{2}\fr{g^{00}g^{11}g^{22}}{\alpha+\beta G^3_3}. \label{coupling}
    \ee
Therefore, the Lagrangian~\eqref{Lag} takes the following form:
    \begin{align}
    L_{\rm 4D}&=\Mpl^2\fr{a^3}{N}\bra{-3\fr{\dot{a}^2}{a^2}+3\dot{\si}^2-\Lambda N^2}
    +\fr{e^{-4\si}\dot{v}^2}{2Na}
    \brc{\alpha-\fr{\beta}{N^2}\brb{\fr{\dot{a}^2}{a^2}+6\fr{\dot{a}}{a}\dot{\si}+3\dot{\si}^2+2\fr{\ddot{a}}{a}+2\ddot{\si}-2\fr{\dot{N}}{N}\bra{\fr{\dot{a}}{a}+\dot{\si}}}}^{-1}, \label{miniLag4D}
    \end{align}
where we have performed an integration by parts.
It should be noted that the set of equations of motion~(EOMs) obtained from the minisuperspace Lagrangian~\eqref{miniLag4D} reproduces the full set of EOMs derived from the original Lagrangian~\eqref{Lag}, and thus the former properly describes the dynamics of the metric~\eqref{gmn4D} and the 2-form field~\eqref{Bmn4D}.
For a given Lagrangian~$L=L(N,\dot{N},a,\dot{a},\ddot{a},\si,\dot{\si},\ddot{\si},\dot{v})$, we write the EL equations as
    \be
    \begin{array}{ll}
    \displaystyle \mE_N \coloneqq \fr{\pa L}{\pa N}-\fr{d}{dt}\bra{\fr{\pa L}{\pa \dot{N}}}=0,
    &\displaystyle \mE_a \coloneqq \fr{\pa L}{\pa a}-\fr{d}{dt}\bra{\fr{\pa L}{\pa \dot{a}}}+\fr{d^2}{dt^2}\bra{\fr{\pa L}{\pa \ddot{a}}}=0, \\
    &\\[-7.5pt]
    \displaystyle \mE_\si \coloneqq \fr{\pa L}{\pa \si}-\fr{d}{dt}\bra{\fr{\pa L}{\pa \dot{\si}}}+\fr{d^2}{dt^2}\bra{\fr{\pa L}{\pa \ddot{\si}}}=0, \quad
    &\displaystyle \mE_v \coloneqq -\fr{d}{dt}\bra{\fr{\pa L}{\pa \dot{v}}}=0.
    \end{array}
    \ee
Due to the presence of second derivatives in the Lagrangian, each EL equation contains third or fourth derivatives.
Note that, in the present case where the Lagrangian is invariant under time diffeomorphisms, not all these equations are independent as the following relation exists:
    \be
    N\dot{\mE}_N-\dot{a}\mE_a-\dot{\si}\mE_\si-\dot{v}\mE_v=0, \label{Noether}
    \ee
which is nothing but the Noether identity associated with the time diffeomorphism symmetry~(see Ref.~\cite{Motohashi:2016prk} for related discussions).
Hence, for instance, $\mE_a=0$ automatically follows from the other three EL equations, and thus one can focus on $\mE_N=0$, $\mE_\si=0$, and $\mE_v=0$.
Having derived the EOMs, one can now use the time diffeomorphism symmetry to fix $N=1$.
It should be noted that this gauge fixing is not complete and one can still freely choose the time origin, $t\to t+t_0$, with $t_0$ being constant.
The EL equation for $v$ can be immediately integrated to give
    \be
    \fr{\pa L}{\pa \dot{v}}=
    \fr{e^{-4\si}\dot{v}}{a}
    \brb{\alpha-\beta\bra{\fr{\dot{a}^2}{a^2}+6\fr{\dot{a}}{a}\dot{\si}+3\dot{\si}^2+2\fr{\ddot{a}}{a}+2\ddot{\si}}}^{-1}
    ={\rm const}\eqqcolon K. \label{EOMv4D}
    \ee
Using this fact, one can eliminate $v$ from $\mE_N$ and $\mE_\si$, which simultaneously removes higher derivatives of $a$ and $\si$ as follows:
    \begin{align}
    \mE_N&=\Mpl^2\,a^3\bra{3\fr{\dot{a}^2}{a^2}-3\dot{\si}^2-\Lambda}
    -\fr{K^2ae^{4\si}}{2}\brb{\alpha+\beta\bra{\fr{\dot{a}^2}{a^2}-4\fr{\dot{a}}{a}\dot{\si}-5\dot{\si}^2}}, \label{EOMN4D} \\
    \mE_\si&=-6\Mpl^2\,a^3\bra{\ddot{\si}+3\fr{\dot{a}}{a}\dot{\si}}
    +K^2ae^{4\si}\brb{-2\alpha+\beta\bra{2\fr{\dot{a}^2}{a^2}+5\fr{\dot{a}}{a}\dot{\si}+10\dot{\si}^2+2\fr{\ddot{a}}{a}+5\ddot{\si}}}. \label{EOMsi4D}
    \end{align}
Now let us count the number of physical DOFs.
From Eqs.~\eqref{EOMv4D}--\eqref{EOMsi4D}, the dynamics of $a$, $\si$, and $v$ can be fixed up to five integration constants, where we have taken into account $K$ in Eq.~\eqref{EOMv4D}.
Since one among the five integration constants can be absorbed into a shift of time~$t\to t+t_0$, the number of physical DOFs of the system is $(5-1)/2=2$.
Notably, the same discussion works for the case~$\beta=0$, showing that the model~\eqref{Lag} with $\beta\ne 0$ does not yield extra DOFs in the minisuperspace description.
To emphasize, it was crucial here that the removal of $v$ from $\mE_N$ and $\mE_\si$ also removes higher derivatives of $a$ and $\si$ at the same time.

\section{Apparent ghost under truncation} \label{sec:truncation}
In the previous section, we saw that the minisuperspace Lagrangian~\eqref{Lag} does not yield unwanted DOFs. 
Actually, this feature is realized only through the nonperturbative interaction manifested by $(\alpha+\beta G^3_3)^{-1}$ in Eq.~\eqref{coupling}.
To see this, let us expand Eq.~\eqref{miniLag4D} with respect to $\beta$ and consider the following Lagrangian truncated at the linear order in $\beta$:
    \be
    L_{\rm lin}=\Mpl^2\fr{a^3}{N}\bra{-3\fr{\dot{a}^2}{a^2}+3\dot{\si}^2-\Lambda N^2}
    +\fr{e^{-4\si}\dot{v}^2}{2\alpha^2Na}
    \brc{\alpha+\fr{\beta}{N^2}\brb{\fr{\dot{a}^2}{a^2}+6\fr{\dot{a}}{a}\dot{\si}+3\dot{\si}^2+2\fr{\ddot{a}}{a}+2\ddot{\si}-2\fr{\dot{N}}{N}\bra{\fr{\dot{a}}{a}+\dot{\si}}}}. \label{tLag}
    \ee
As we did in Eq.~\eqref{EOMv4D}, the EL equation for $v$ derived from this Lagrangian can be integrated to yield $\dot{v}$ as a function of $a$, $\si$, and their derivatives,
\begin{align}
 \dot{v} = \alpha Kae^{4\si}\left[ 1 + \frac{\beta}{\alpha} \left(
 \frac{\dot{a}^2}{a^2} + 6 \frac{\dot{a}}{a} \dot{\sigma} + 3 \dot{\sigma}^2 + 2 \frac{\ddot{a}}{a} + 2 \ddot{\sigma} 
\right) \right]^{-1},
\end{align}
where $K$ being an integration constant.
Note that here we have set $N=1$.
After eliminating $v$ from the EL equations for $N$ and $\si$, there still remain higher derivatives,
    \begin{align}
    \mE_N&=-\fr{4\beta^2K^2a^7e^{4\si}}{\alpha^5}\left[ 1+\fr{\beta}{\alpha}\left(
 \frac{\dot{a}^2}{a^2} + 6 \frac{\dot{a}}{a} \dot{\sigma} + 3 \dot{\sigma}^2 + 2 \frac{\ddot{a}}{a} + 2 \ddot{\sigma} 
\right) \right]^{-3}\bra{\fr{\dot{a}}{a}+\dot{\si}}\bra{\fr{\dddot{a}}{a}+\dddot{\si}}+\cdots, \nonumber \\
    \mE_\si&=-\fr{4\beta^2K^2a^7e^{4\si}}{\alpha^5}\left[ 1+\fr{\beta}{\alpha}\left(
 \frac{\dot{a}^2}{a^2} + 6 \frac{\dot{a}}{a} \dot{\sigma} + 3 \dot{\sigma}^2 + 2 \frac{\ddot{a}}{a} + 2 \ddot{\sigma} 
\right) \right]^{-3}\bra{\fr{\ddddot{a}}{a}+\ddddot{\si}}+\cdots,
    \end{align}
where the ellipses denote terms with lower derivatives.
This is in sharp contrast to Eqs.~\eqref{EOMN4D} and \eqref{EOMsi4D} in the previous section.
From the above expressions, one can see that the fourth derivative terms in $\mE_\si$ can be eliminated by taking the following linear combination with $\mE_N$:
    \be
    \mE_\si-\fr{a}{\dot{a}+a\dot{\si}}\dot{\mE}_N.
    \ee
Notably, after taking this linear combination, the third derivative terms are also removed. 
Thus, we are left with one third-order and two second-order differential equations, namely, 
    \be
    \mE_N=0,\quad
    \mE_v=0,\quad
    \mE_\si-\fr{a}{\dot{a}+a\dot{\si}}\dot{\mE}_N=0.
    \ee
Hence, the number of DOFs of the system is $(7-1)/2=3$, where one among the seven integration constants has been absorbed into a shift of time.
This implies that the truncated Lagrangian~\eqref{tLag} has one extra DOF compared to the Lagrangian~\eqref{miniLag4D}.
This is consistent with the no-go result in Ref.~\cite{Heisenberg:2019akx}. 
The important point here is that no higher derivatives appear up to ${\cal O}(\beta)$. 
Such higher derivatives are possibly canceled out by adding new derivative interactions of ${\cal O}(\beta^2)$ to the truncated Lagrangian~\eqref{tLag}.
By repeating this procedure, one could construct a ghost-free theory with an infinite number of coupling terms. 
Our model~\eqref{miniLag4D} can be understood as an example of such a nonperturbative completion.

\section{Ghost in higher dimensions}\label{sec:higherD}

Having established the absence of ghost in four dimensions, let us now consider a $D$-dimensional spacetime with $D\ge 5$,
    \be
    L_D=\sqrt{-g}\brb{\fr{\Mpl^{D-2}}{2}(R-2\Lambda)-\fr{1}{12}\fr{1}{\det\mG}\mG^{\mu\alpha}\mG^{\nu\beta}\mG^{\la\ga}H_{\mu\nu\la}H_{\alpha\beta\ga}}. \label{LagD}
    \ee
This Lagrangian is a natural extension of Eq.~\eqref{Lag}.
However, the duality to a healthy scalar-tensor theory in four dimensions no longer holds in higher dimensions, and thus it is nontrivial whether the model~\eqref{LagD} has ghostly DOFs.
In this section, we address this issue in a minisuperspace description.
We assume the metric and the 2-form field of the following form:
    \begin{align}
    &g_\mn dx^\mu dx^\nu=-N^2(t)dt^2+a^2(t)\brb{e^{2\si(t)}\Bigl((dx^1)^2+(dx^2)^2\Bigr)+e^{-\fr{4\si(t)}{D-3}}\sum_{i=3}^{D-1}(dx^i)^2}, \label{gmnD} \\
    &\fr{1}{2}B_\mn dx^\mu \wedge dx^\nu=v(t)dx^1 \wedge dx^2, \label{BmnD}
    \end{align}
which is a generalization of Eqs.~\eqref{Bmn4D} and \eqref{gmn4D}.
As in the previous section, the coupling between the 2-form field and the curvature tensor is simplified as
    \be
    -\fr{1}{12}\fr{1}{\det\mG}\mG^{\mu\alpha}\mG^{\nu\beta}\mG^{\la\ga}H_{\mu\nu\la}H_{\alpha\beta\ga}
    =-\fr{\dot{v}^2}{2}\fr{\mG^{00}\mG^{11}\mG^{22}}{\mG^0_0\mG^1_1\mG^2_2\mG^3_3\cdots\mG^{D-1}_{D-1}}
    =-\fr{\dot{v}^2}{2}\fr{g^{00}g^{11}g^{22}}{(\mG^3_3)^{D-3}}.
    \ee
Here, we have used the fact that $\mG^3_3=\cdots=\mG^{D-1}_{D-1}$ and $\mG^{3}_{3}$ is given by
\begin{align}
 \mG^{3}_{3} = \alpha-\fr{\beta}{N^2}\Biggl[&\fr{(D-2)(D-3)}{2}\fr{\dot{a}^2}{a^2}+2(D-1)\fr{\dot{a}}{a}\dot{\si}+\fr{D-1}{D-3}\dot{\si}^2 \nonumber \\
    &+(D-2)\fr{\ddot{a}}{a}+\fr{2}{D-3}\ddot{\si}-\fr{\dot{N}}{N}\bra{(D-2)\fr{\dot{a}}{a}+\fr{2}{D-3}\dot{\si}}\Biggr].
\end{align}
Thus, the Lagrangian~\eqref{LagD} is written as
    \begin{align}
    L_{D}=\Mpl^{D-2} \fr{a^{D-1}}{N}\brb{-\fr{(D-1)(D-2)}{2}\fr{\dot{a}^2}{a^2}+\fr{D-1}{D-3}\dot{\si}^2-\Lambda N^2}+\fr{a^{D-5}e^{-4\si}\dot{v}^2}{2N(\mG^3_3)^{D-3}}, \label{miniLagD}
    \end{align}
where we have performed an integration by parts.
Note that Eq.~\eqref{miniLagD} reduces to Eq.~\eqref{miniLag4D} when $D=4$.
Note also that this minisuperspace Lagrangian properly describes the dynamics of the metric~\eqref{gmnD} and the 2-form field~\eqref{BmnD} in the sense that it yields a set of EOMs which is equivalent to the one obtained from the original Lagrangian~\eqref{Lag}.
One notices that the structure of the Lagrangian~\eqref{miniLagD} in higher dimensions is similar to that of Eq.~\eqref{miniLag4D} in four dimensions, except for the exponent of $\mG^3_3(=\alpha+\beta G^3_3)$.
Now we study the structure of independent EOMs~$\mE_N=0$, $\mE_\si=0$, and $\mE_v=0$.
In what follows, we put $N=1$.
As in Eq.~\eqref{EOMv4D}, the EOM for $v$ can be integrated to yield
\begin{align}
\dot{v} = K a^{5-D}e^{4\si} \left[ 2(D-3) \mG^{3}_{3} \right]^{D-3},
\end{align}
with $K$ being an integration constant.
Plugging this equation into $\mathcal{E}_{N}$ and $\mathcal{E}_{\sigma}$, we obtain
\begin{align}
 \mathcal{E}_{N} &= -A \left( \frac{(D-2)(D-3)}{2} \frac{\dot{a}}{a} + \dot{\sigma} \right) \left( \frac{(D-2)(D-3)}{2} \frac{\dddot{a}}{a} + \dddot{\sigma}\right) +\cdots, \\
 \mathcal{E}_{\sigma} &= -A \left( \frac{(D-2)(D-3)}{2} \frac{\ddddot{a}}{a} + \ddddot{\sigma}\right)+\cdots, 
\end{align}
where the ellipses denote lower derivative terms.
Here, we have defined the coefficient~$A$ as
    \be
    A\coloneqq 2^{2D-5} (D - 4) ( D - 3 )^{2D-7} \beta^2 K^2 a^{5-D} e^{4 \sigma}   (\mG^{3}_{3})^{D-5}.
    \ee
Note that $A$ contains a factor~$(D-4)$, which reflects the absence of higher derivative terms in Eqs.~\eqref{EOMN4D} and \eqref{EOMsi4D} in the case of four dimensions.
Thus, one can remove $\ddddot{a}$ and $\ddddot{\si}$ by taking a linear combination of $\mE_\si$ and $\dot{\mE}_N$.
After this manipulation, we have
    \begin{align}
    \mE_\si-\fr{2a}{(D-2)(D-3)\dot{a}+2a\dot{\si}}\dot{\mE}_N=A \fr{(D-4)(D-1)^2\dot{a}\dot{\si}}{(D-2)(D-3)\dot{a}+2a\dot{\si}} \left( \frac{(D-2)(D-3)}{2} \frac{\dddot{a}}{a} + \dddot{\sigma}\right)+\cdots,
    \end{align}
where the third derivative terms can be eliminated by using $\mE_N=0$.
We are now left with one third-order and two second-order differential equations,
    \be
    \mE_N=0,\quad
    \mE_v=0,\quad
    \mE_\si-\fr{(D-4)(D-1)^2\dot{a}\dot{\si}}{(D-2)(D-3)\dot{a}+2a\dot{\si}}\dot{\mE}_N+\fr{2(D-4)(D-1)^2a\dot{a}\dot{\si}}{\brb{(D-2)(D-3)\dot{a}+2a\dot{\si}}^2}\mE_N=0.
    \ee
Recalling that there is a remaining gauge DOF corresponding to the shift of time origin, we need six initial conditions to fix the dynamics of $a$, $\si$, and $v$.
This means that the system has three DOFs and thus we have one unwanted extra DOF.

\section{Conclusions and discussion}\label{sec:conc}

In the present paper, we developed a minisuperspace description of the model~\eqref{Lag} and clarified how the nonperturbative higher derivative structure helps to circumvent the problem of the Ostrogradsky ghost.
The key feature is that the EOM for the 2-form gauge field can be integrated to give the relation~\eqref{EOMv4D}, which miraculously cancels all the higher derivative terms in the other EOMs.
As we saw in \S \ref{sec:truncation}, this cancellation no longer occurs once the Lagrangian is truncated at a finite order.
It was also shown in \S \ref{sec:higherD} that, when proceeding to higher dimensions, this cancellation does not occur, and thus the model~\eqref{Lag} is healthy only if it is defined in four dimensions.

The minisuperspace Lagrangian~\eqref{miniLag4D} is structurally similar to the toy example studied in Sec.~V\hspace{-.1em}I\hspace{-.1em}I~A of Ref.~\cite{Gabadadze:2012tr}, which is of the form
    \be
    L^{\rm toy}=\fr{1}{2}\dot{\phi}^2+\fr{1}{2}\fr{1}{1+\beta\ddot{\phi}}\dot{\psi}^2, \label{Ltoy}
    \ee
with $\phi$ and $\psi$ being functions of $t$ and $\beta$ being a nonvanishing constant.
Indeed, both Lagrangians~\eqref{miniLag4D} and \eqref{Ltoy} have second derivative(s) in the denominator.
The toy model~\eqref{Ltoy} elucidates the reason why the system~\eqref{miniLag4D} is free of unwanted extra DOFs.
One may first observe that the Lagrangian~\eqref{Ltoy} would yield extra DOFs compared to a system of two point particles due to the factor~$(1+\beta\ddot{\phi})^{-1}$ in front of $\dot{\psi}^2$.
However, this is not the case as shown below.
From the Lagrangian~\eqref{Ltoy}, we obtain the EL equations as
    \begin{align}
    \ddot{\phi}+\fr{\beta}{2}\fr{d^2}{dt^2}\brb{\bra{\fr{\dot{\psi}}{1+\beta\ddot{\phi}}}^2}=0, \quad
    \fr{d}{dt}\bra{\fr{\dot{\psi}}{1+\beta\ddot{\phi}}}=0.
    \end{align}
We have $\dot{\psi}/(1+\beta\ddot{\phi})={\rm const}$ from the second equation, which can be substituted into the first equation to yield $\ddot{\phi}=0$.
Then, resubstituting this into the second equation, we obtain $\ddot{\psi}=0$.
Hence, we need four initial conditions to specify the solution for $\phi$ and $\psi$.
This means that the system described by the Lagrangian~\eqref{Ltoy} has two DOFs, which is the same number of DOFs as a system of two point particles (i.e., the case~$\beta=0$) has.

The situation changes if one expands the Lagrangian with respect to $\beta$ and truncates at the linear order,
    \be
    L^{\rm toy}_{\rm lin}=\fr{1}{2}\dot{\phi}^2+\fr{1}{2}(1-\beta\ddot{\phi})\dot{\psi}^2. \label{Ltoylin}
    \ee
The EL equations are as follows:
    \be
    \ddot{\phi}+\fr{\beta}{2}\fr{d^2}{dt^2}\bra{\dot{\psi}^2}=0, \quad
    \fr{d}{dt}\brb{(1-\beta\ddot{\phi})\dot{\psi}}=0.
    \ee
As in the previous case of $L^{\rm toy}$, one can integrate the second equation to obtain $(1-\beta\ddot{\phi})\dot{\psi}={\rm const}\eqqcolon c$.
Then, one can eliminate $\dot{\psi}$ from the first equation to yield a fourth-order differential equation for $\phi$,
    \be
    \ddot{\phi}+\fr{c^2\beta}{2}\fr{d^2}{dt^2}\brb{\fr{1}{(1+\beta\ddot{\phi})^{2}}}=0.
    \ee
Thus, this system needs two more initial conditions than the system described by Eq.~\eqref{Ltoy}.
This is similar to what we saw in \S \ref{sec:4D}:
By linearizing the Lagrangian~\eqref{miniLag4D} with respect to $\beta$, we found that additional initial conditions are needed to fully solve the system of EOMs.

It is also possible to understand the result of \S \ref{sec:higherD} through the following Lagrangian:
    \be
    L^{\rm toy}_{n}=\fr{1}{2}\dot{\phi}^2+\fr{1}{2}\fr{1}{(1+\beta\ddot{\phi})^n}\dot{\psi}^2, \label{Ltoyn}
    \ee
with $n\ge 2$ being an integer.
Note that now the exponent of $(1+\beta\ddot{\phi})$ is different from $-1$, which is the same situation we encountered in Eq.~\eqref{miniLagD}.
The EL equations read
    \begin{align}
    \ddot{\phi}+\fr{n\beta}{2}\fr{d^2}{dt^2}\brb{\fr{\dot{\psi}^2}{(1+\beta\ddot{\phi})^{n+1}}}=0, \quad
    \fr{d}{dt}\brb{\fr{\dot{\psi}}{(1+\beta\ddot{\phi})^{n}}}=0.
    \end{align}
From the second equation, we obtain $\dot{\psi}/(1+\beta\ddot{\phi})^n={\rm const}\eqqcolon c$.
Using this, one can remove $\dot{\psi}$ from the first equation to yield a fourth-order differential equation for $\phi$,
    \be
    \ddot{\phi}+\fr{c^2n\beta}{2}\fr{d^2}{dt^2}\brb{(1+\beta\ddot{\phi})^{n-1}}=0.
    \ee
Thus, compared to the model~\eqref{Ltoy}, two more initial conditions are necessary to fix the dynamics of $\phi$ and $\psi$, meaning the presence of an extra DOF.
This corresponds to the appearance of one extra DOF in the system~\eqref{miniLagD}.

So far, we have clarified the ghost-free structure of the model~\eqref{Lag} under the minisuperspace description. 
It would be interesting to check if the same structure holds in the general Arnowitt-Deser-Misner formalism without using the minisuperspace ansatz.
Another thing of interest is to explore a ghost-free resummation for a field other than 2-form, e.g., a scalar or vector field. 
We leave these issues for future work.


\acknowledgments{
We would like to thank Jiro Soda for enlightening discussions.
K.T.\ is supported by JSPS KAKENHI Grants No.\ JP17H02894 and No.\ JP17K18778.
D.Y.\ is supported by the JSPS Postdoctoral Fellowship No.\ 201900294.
We are also supported by JSPS Bilateral Joint Research
Projects (JSPS-NRF Collaboration) ``String Axion Cosmology.''


\bibliographystyle{mybibstyle}
\bibliography{2-form}

\end{document}